\documentclass[dvips,12pt,a4paper]{article}
\usepackage{a4wide}
\usepackage{epsfig}
\usepackage{amsmath}
\usepackage{latexsym}

  \newlength{\absize}
\newcommand{\dd}{\mbox{{\rm d}}}

\newcommand{\Lumint}{{\cal L}_{\rm int}}

\catcode`@=11
\def\citer{\@ifnextchar [{\@tempswatrue\@citexr}{\@tempswafalse\@citexr[]}}
 
\def\@citexr[#1]#2{\if@filesw\immediate\write\@auxout{\string\citation{#2}}\fi
  \def\@citea{}\@cite{\@for\@citeb:=#2\do
    {\@citea\def\@citea{--\penalty\@m}\@ifundefined
       {b@\@citeb}{{\bf ?}\@warning
       {Citation `\@citeb' on page \thepage \space undefined}}%
\hbox{\csname b@\@citeb\endcsname}}}{#1}}
\catcode`@=12

\begin{document}
  \thispagestyle{empty}
  \pagestyle{empty}
  \renewcommand{\thefootnote}{\fnsymbol{footnote}}
\newpage\normalsize
    \pagestyle{plain}
    \setlength{\baselineskip}{4ex}\par
    \setcounter{footnote}{0}
    \renewcommand{\thefootnote}{\arabic{footnote}}
\newcommand{\preprint}[1]{%
  \begin{flushright}
    \setlength{\baselineskip}{3ex} #1
  \end{flushright}}
\renewcommand{\title}[1]{%
  \begin{center}
    \LARGE #1
  \end{center}\par}
\renewcommand{\author}[1]{%
  \vspace{2ex}
  {\Large
   \begin{center}
     \setlength{\baselineskip}{3ex} #1 \par
   \end{center}}}
\renewcommand{\thanks}[1]{\footnote{#1}}
\renewcommand{\abstract}[1]{%
  \vspace{2ex}
  \normalsize
  \begin{center}
    \centerline{\bf Abstract}\par
    \vspace{2ex}
    \parbox{\absize}{#1\setlength{\baselineskip}{2.5ex}\par}
  \end{center}}

\begin{flushright}
%{\setlength{\baselineskip}{2ex}\par
%{\tt University of Bergen, Department of Physics}    \\[1mm]
%{\tt Scientific/Technical Report No.\ 1999-03}    \\[1mm]
%{\tt ISSN 0803-2696} \\[2mm]
{hep-ph/9910403} \\
%[1mm]
{October 1999}           \\
\end{flushright}
\vspace*{4mm}
\vfill
\title{Optimal polarized observables for model-independent 
new-physics search at $e^+e^-$  colliders}
\vfill
\author{
A.A. Babich$^{a}$,
P. Osland$^{b}$,
A.A. Pankov$^{a,c}$, {\rm and}
N. Paver$^{c}$}
%-----------------------------------
%   Address
%-----------------------------------
%\vspace{1cm}
\begin{center}
$^a$ Pavel Sukhoi Technical University, 
     Gomel, 246746 Belarus \\
$^b$ Department of Physics, University of Bergen, \\
     All\'{e}gaten 55, N-5007 Bergen, Norway \\
$^c$ Dipartimento di Fisica Teorica, Universit\`a di Trieste and \\
Istituto Nazionale di Fisica Nucleare, Sezione di Trieste,
Trieste, Italy
\end{center}
\vfill
\abstract
{For the processes $e^+e^-\to \mu^+\mu^-$, $\tau^+\tau^-$, $b\bar{b}$ 
and $c\bar{c}$ at a future $e^+e^-$ 
collider with $\sqrt{s}=0.5$ TeV, we examine the sensitivity of the helicity
cross sections to four-fermion contact interactions. If longitudinal 
polarization of the electron beam were available, two polarized integrated 
cross sections would offer the opportunity to separate the helicity cross 
sections and, in this way, to derive model-independent bounds on the relevant 
parameters. The measurement of these polarized cross sections with optimal 
kinematical cuts could significantly increase the sensitivity of helicity 
cross sections to contact interaction parameters and could give crucial 
information on the chiral structure of such new interactions.
}
\vspace*{20mm}
\setcounter{footnote}{0}
\vfill

\newpage
    \setcounter{footnote}{0}
    \renewcommand{\thefootnote}{\arabic{footnote}}
    \setcounter{page}{1}
%%%%%%%%%%%%%%%%%%%%%%%%%%%%%%%%%%%%%%%%%%%%%%%%%%%%%%%%%%%%%%%%%%%%%%%%
\section{Introduction}
%%%%%%%%%%%%%%%%%%%%%%%%%%%%%%%%%%%%%%%%%%%%%%%%%%%%%%%%%%%%%%%%%%%%%%%%

Any measured deviation from the Standard Model (SM) predictions in
electron-positron annihilation into fermion-pairs
\begin{equation}
e^++e^-\to f+\bar{f} \label{proc}
\end{equation}
%%\rightline{proc}
($f=l$ or $q$ for lepton or quark) would signal the presence of new phenomena. 
By means of the effective Lagrangian approach, four-fermion contact 
interactions offer a general framework for describing low-energy 
manifestations of non-standard interactions active at a very high energy scale 
$\Lambda$, which in many cases can be interpreted as the mass of a new heavy 
particle exchanged in the process, with a strength governed by some effective 
coupling constant $g_{\rm eff}$ \cite{Eichten, Ruckl}. 

At sub-TeV energies, representative examples are the exchange of a 
$Z^\prime$ with a few TeV mass and the exchange of a heavy 
leptoquark. In general, any new interaction generated by $s$, $t$ or $u$ 
exchanges of objects with mass-squared much larger than the corresponding 
Mandelstam variables can be described by effective four-fermion $eeff$ local 
interactions \cite{Schrempp,Barger2}. This is also the case,  
in the context of supersymmetry, of $R$-parity breaking interactions 
at energies 
much smaller than the sparticle masses, and of the composite models of quarks 
and leptons, where contact interactions arise as a remnant of the binding 
force between the fermion substructure constituents. 

Thus, the concept of contact interactions with a universal energy scale 
$\Lambda$ is quite generally used, also in other processes besides 
(\ref{proc}) such as $ep$ and $p\bar{p}$ collisions, to conveniently 
parameterize deviations from the SM that may be caused by some new physics at 
the large energy scale~$\Lambda$. 

The lowest-order four-fermion contact terms have dimension $D=6$, 
which implies 
that they are suppressed by $g^2_{\rm eff}/\Lambda^2$. Restricting the fermion 
currents to be helicity conserving and flavor diagonal, the general 
$SU(3)\times SU(2)\times U(1)$ invariant four-fermion $eeff$ contact 
interaction Lagrangian with $D=6$ can be written as 
\citer{Eichten,Barger2}:
\begin{eqnarray}
{{\cal L}}=\frac{g^2_{\rm eff}}{\Lambda^2}\left[
\eta_{\rm LL}\left(\bar e_{\rm L}\gamma_\mu e_{\rm L}\right)
\left(\bar f_{\rm L}\gamma^\mu f_{\rm L}\right) 
+\eta_{\rm LR}\left(\bar{e}_{\rm L}\gamma_\mu e_{\rm L}\right)
\left(\bar f_{\rm R}\gamma^\mu f_{\rm R}
\right)\right. \nonumber \\
+\left.\eta_{\rm RL}\left(\bar{e}_{\rm R}\gamma_\mu e_{\rm R}\right)
\left(\bar f_{\rm L}
\gamma^\mu f_{\rm L}\right) +
\eta_{\rm RR}\left(\bar e_{\rm R}\gamma_\mu e_{\rm R}\right)
\left(\bar f_{\rm R}\gamma^\mu f_{\rm R}\right) 
\right],
\label{lagra}
\end{eqnarray}
%%\rightline{lagra}
where generation and color indices have been suppressed.\footnote{For the case
of ${\bar t}t$ final states, see \cite{Grzadkowski}.} The subscripts 
L and R indicate that the current in each parenthesis can be either left- or 
right-handed, and the parameters $\eta_{\alpha\beta}$ ($\alpha,\beta=$ L,
R) determine the chiral structure of the interaction. 
Although in a purely phenomenological approach they are free parameters, 
conventionally they are taken to be $\eta_{\alpha\beta}=\pm 1\ {\rm or}\ 0$.
%$-1$ or $+1$, depending on the type of the assumed theory \cite{Schrempp}. 
Also, it is conventional to take 
$g^2_{\rm eff}=4\pi$, as reminiscence of the fact that such contact 
interactions were initially introduced in the framework of compositeness, 
where the new binding forces were assumed to be strong.

In general, for a given fermion flavor $f$, Eq.~(\ref{lagra}) defines eight 
independent, individual interaction models corresponding to the combinations 
of the four chiralities LL, LR, RL and RR with the $\pm$ signs of the 
$\eta$'s. In practice, the true interaction might correspond to one of these 
models or to any combination of them. The number of independent
coefficients could be reduced by imposing symmetries that provide relations 
among the contact interaction couplings \cite{Barger2}.\footnote{For
example, for $Z'$ exchange, $\eta_{\rm LL}\eta_{\rm RR}
=\eta_{\rm LR}\eta_{\rm RL}$.}
Here, we will not consider any particular symmetry.
We will instead perform a model-independent analysis considering 
contact interaction couplings of magnitude $\eta_{\alpha\beta}=\pm 1$, 
and independent mass-scales $\Lambda_{\alpha\beta}$ corresponding 
to each of the helicity combinations in Eq.~(\ref{lagra}). 
For our purpose of obtaining constraints on the $\Lambda_{\alpha\beta}$,
the signs of the $\eta$'s turn out to be numerically unimportant. 
Indeed, for given helicities $\alpha\beta$, different signs $\eta$ yield 
practically identical results for the mass scales 
$\Lambda_{\alpha\beta}$ because, in the chosen kinematical configuration, 
the non-standard effects are largely dominated by the interference 
between the contact interaction and the SM terms.

Clearly, the definition of $\Lambda$ adopted here,
provides a standard for comparing the power of different
new-physics searches.
For example, a bound on $\Lambda$ would mean that a $Z'$ with couplings 
of the order of the electromagnetic couplings could exist with
mass down to $M_{Z^\prime}\sim\sqrt{\alpha}\,\Lambda$, and the same
would be true for leptoquarks and for any new, very heavy gauge bosons
that may be exchanged in the process under consideration.

In principle, the sought-for deviations of observables from the SM predictions,
giving information on $\Lambda$'s, simultaneously depend on all four-fermion 
effective coupling constants in Eq.~(\ref{lagra}), which therefore cannot be 
easily disentangled. For simplicity, the analysis is usually performed by 
taking a non-zero value for only one parameter at a time, all the remaining 
ones being put equal to zero. Limits on individual $eeqq$ contact 
interaction parameters have recently been derived by this procedure, 
from a global analysis of the relevant data \cite{Barger2}, and  
the individual models are severely constrained, with 
$\Lambda_{\alpha\beta}\sim{\cal{O}}(10)$ TeV. 

However, if several terms of different chiralities were simultaneously taken
into account, cancellations may occur and the resulting bounds on
$\Lambda_{\alpha\beta}$ would be considerably weaker, of the order of 
$3-4$ TeV. As an example, the constraints originating from atomic parity 
violation experiments could be substantially relaxed if there were 
compensating contributions from different terms \cite{Barger3}.
Consequently, a definite improvement of the situation in this regard should be
obtained from a procedure of analyzing experimental data that 
allows to account for the various contact interaction couplings 
simultaneously as 
free parameters, and yet to obtain in a model-independent way separate bounds 
for the corresponding $\Lambda$'s, not affected by possible accidental 
cancellations.    

For that purpose, we propose an analysis of $eell$, $eebb$ 
and $eecc$ contact interactions at the next linear $e^+e^-$ collider (LC) with 
$\sqrt s= 500$ GeV and with longitudinally polarized beams.
Our approach makes use of two particular, polarized, integrated cross sections 
$\sigma_1$ and $\sigma_2$ (see Eqs.~(\ref{sigma1}) and (\ref{sigma2})), that  
are directly connected, {\it via} linear combinations, to the helicity cross 
sections of process (\ref{proc}), and therefore allow to deal with a minimal 
set of independent free parameters.  

This kind of observables, defined for specific kinematical cuts, were already 
introduced to study $Z^\prime$ signals at LEP2 and LC \cite{Osland,Babich} and 
potential manifestations of compositeness at the LC \cite{Pankov}. Here, we 
extend the previous considerations by performing a general analysis where,  
in the definition of the above-mentioned integrated observables, we choose 
suitable kinematical regions where the sensitivity to individual four-fermion 
contact interaction parameters is maximal. 
As we shall see, this procedure of optimization 
results in a further increase of sensitivity, that for some of the 
four-fermion interactions can be quite substantial. Moreover, we make 
a short comparison of the numerical bounds on $\Lambda$'s with the results
of the more commonly used observables.

%%%%%%%%%%%%%%%%%%%%%%%%%%%%%%%%%%%%%%%%%%%%%%%%%%%%%%%%%%%%%%%%%%%%%%%%
\section{Polarized observables}
%%%%%%%%%%%%%%%%%%%%%%%%%%%%%%%%%%%%%%%%%%%%%%%%%%%%%%%%%%%%%%%%%%%%%%%%
In the Born approximation, including the $\gamma$ and $Z$ exchanges
as well as the four-fermion
contact interaction term (\ref{lagra}), but neglecting $m_f$ with respect
to the c.m.\ energy $\sqrt s$, the differential cross section for the process
$e^+e^-\to  f\bar{f}$ ($f\neq e, t$)
with longitudinally polarized electron-positron beams,
can be written as  \cite{Zeppenfeld2}
\begin{equation}
\frac{\dd\sigma}{\dd\cos\theta}
=\frac{3}{8}
\left[(1+\cos\theta)^2 {\sigma}_+
+(1-\cos\theta)^2 {\sigma}_-\right],
\label{cross}
\end{equation}
%%%\rightline{cross}
where $\theta$ is the angle between the incoming electron and the outgoing
fermion in the c.m.\ frame.
The functions $\sigma_\pm$ can be expressed in terms of helicity
cross sections
\begin{equation}
\sigma_{\alpha\beta}=N_C\sigma_{\rm pt}
\vert A_{\alpha\beta}\vert^2,
\label{helcross}
\end{equation}
with $\alpha,\beta={\rm L,R}$.
Here, $N_C$ is the QCD factor: $N_C\approx 3(1+\alpha_s/\pi)$
for quarks and  $N_C=1$ for leptons, respectively, and
$\sigma_{\rm pt}\equiv\sigma(e^+e^-\to\gamma^\ast\to l^+l^-)
=(4\pi\alpha^2)/(3s)$.
With electron and positron longitudinal polarizations $P_e$ and $P_{\bar e}$,
the relations are
\begin{eqnarray}
{\sigma}_{+}&=&\frac{1}{4}\,
\left[(1-P_e)(1+P_{\bar{e}})\,\sigma_{\rm LL}
+(1+P_e)(1- P_{\bar{e}})\,\sigma_{\rm RR}\right], \label{s+} \\
{\sigma}_{-}&=&\frac{1}{4}\,
\left[(1-P_e)(1+ P_{\bar{e}})\,\sigma_{\rm LR}
+(1+P_e)(1-P_{\bar{e}})\,\sigma_{\rm RL}\right]. \label{s-}
\end{eqnarray}
The helicity amplitudes
$A_{\alpha\beta}$ can be written as
\begin{equation}
A_{\alpha\beta}=Q_e Q_f+g_\alpha^e\,g_\beta^f\,\chi_Z+
\frac{s\eta_{\alpha\beta}}{\alpha\Lambda_{\alpha\beta}^2},
\label{amplit}
\end{equation}
%%\rightline{amplit}
where the gauge boson propagator is $\chi_Z=s/(s-M^2_Z+iM_Z\Gamma_Z)$,
the SM left- and right-handed fermion couplings of the $Z$ are
$g_{\rm L}^f=(I_{3L}^f-Q_f s_W^2)/s_W c_W$ and
$g_{\rm R}^f=-Q_f s_W^2/s_W c_W$ with
$s_W^2=1-c_W^2\equiv \sin^2\theta_W$, and $Q_f$ are the fermion electric
charges.

The total cross section and the difference of forward and backward cross
sections are given as
\begin{eqnarray}
\label{canon}
\sigma
&=&{\sigma}_{+}+{\sigma}_{-}
=\frac{1}{4}\left[(1-P_e)(1+P_{\bar{e}})(\sigma_{\rm LL}+\sigma_{\rm LR})
+(1+P_e)(1- P_{\bar{e}})(\sigma_{\rm RR}+\sigma_{\rm RL})\right], \\
\sigma_{\rm FB}
&\equiv&\sigma_{\rm F}-\sigma_{\rm B}
=\frac{3}{4}\left({\sigma}_{+}-{\sigma}_{-}\right) \nonumber \\
&=&\frac{3}{16}\left[(1-P_e)(1+P_{\bar{e}})(\sigma_{\rm LL}-\sigma_{\rm LR})
+(1+P_e)(1- P_{\bar{e}})(\sigma_{\rm RR}-\sigma_{\rm RL})\right],
\end{eqnarray}
where
\begin{equation}
\label{sfb}
\sigma_{\rm F}
=\int_{0}^{1}(\dd\sigma/\dd\cos\theta)\dd\cos\theta, \qquad
\sigma_{\rm B}
=\int_{-1}^{0}(\dd\sigma/\dd\cos\theta)\dd\cos\theta.
\end{equation}
Taking Eq.~(\ref{amplit}) into account, these relations show that in general 
$\sigma$ and $\sigma_{\rm FB}$ simultaneously involve all contact-interactions 
couplings even in the polarized case. Therefore, by themselves, these 
measurements do not allow a completely model-independent analysis avoiding, 
in particular, potential cancellations among different couplings. 

Our analysis will be based on the consideration of the four helicity cross 
sections $\sigma_{\alpha\beta}$ as the basic independent observables to be
measured from data on the polarized differential cross section. These cross 
sections depend each on just one individual four-fermion contact parameter 
and therefore lead to a model-independent analysis where all 
$\eta_{\alpha\beta}$ can be taken simultaneously into account as completely 
free parameters, with no danger from potential cancellations. As 
Eqs.~(\ref{s+}) and (\ref{s-}) show, helicity cross sections can be 
disentangled {\it via} the measurement of ${\sigma}_{+}$ and ${\sigma}_{-}$ 
with different choices of the initial beam polarizations.  

One possibility is to project out ${\sigma}_+$ and ${\sigma}_-$ from 
$\dd\sigma/\dd\cos\theta$, as differences of integrated
cross sections. To this aim, we define $z^*_\pm\equiv\cos\theta^*_\pm$ such
that
\begin{equation}
\label{zpm}
\left(\int_{z^*_\pm}^1-\int_{-1}^{z^*_\pm}\right)
\left(1\mp\cos\theta\right)^2\dd\cos\theta=0.
\end{equation}
One finds the solutions:
$z^*_\pm=\mp(2^{2/3}-1)=\mp 0.587$ ($\theta^*_+=126^\circ$ and
$\theta^*_-=54^\circ$).\footnote{These values satisfy
$(z_\pm^*\mp1)^3=\mp4$. In the case of
$\vert\cos\theta\vert=c<1$, one has ${\vert z^*_\pm\vert=(1+3c^2)^{1/3}-1}$.}

From Eq.~(\ref{cross}) one can easily see that at these values of $z^*_\pm$
the difference of two integrated cross sections defined as
\begin{equation}
\label{proj}
\sigma_1(z^*_\pm)-\sigma_2(z^*_\pm)
\equiv \left(\int_{z^*_\pm}^1-\int_{-1}^{z^*_\pm}\right)
\frac{\dd\sigma}{\dd\cos\theta}\,\dd\cos\theta
%=\pm\gamma^{-1}\sigma_\pm,
\end{equation}
is directly related to $\sigma_\pm$ as:
%\begin{eqnarray}
%\label{tildesigma+}
%\sigma_1(z^*_+)-\sigma_2(z^*_+)=\gamma\sigma_{+}, \\
%\sigma_2(z^*_-)-\sigma_1(z^*_-)=\gamma\sigma_{-},
%\label{tildesigma-}
%\end{eqnarray}
\begin{equation}
\label{tildesigma+-}
\sigma_1(z^*_+)-\sigma_2(z^*_+)=\gamma\sigma_{+}, \qquad 
\sigma_2(z^*_-)-\sigma_1(z^*_-)=\gamma\sigma_{-},
\end{equation}
where $\gamma=3\left(2^{2/3}-2^{1/3}\right)=0.982$.

The solutions of the system of two equations corresponding to $P_e=\pm P$,
and assuming  unpolarized positrons $P_{\bar e}=0$,
in Eqs.~(\ref{s+}) and (\ref{s-}), can be written as:
\begin{eqnarray}
\label{SLL}
\sigma_{\rm LL}
&=&\frac{1+P}{P}\sigma_{+}(-P)-\frac{1-P}{P}\sigma_{+}(P), \\
\label{SRR}
\sigma_{\rm RR}
&=&\frac{1+P}{P}\sigma_{+}(P)-\frac{1-P}{P}\sigma_{+}(-P), \\
\label{SLR}
\sigma_{\rm LR}
&=&\frac{1+P}{P}\sigma_{-}(-P)-\frac{1-P}{P}\sigma_{-}(P), \\
\label{SRL}
\sigma_{\rm RL}
&=&\frac{1+P}{P}\sigma_{-}(P)-\frac{1-P}{P}\sigma_{-}(-P).
\end{eqnarray}
From Eqs. (\ref{SLL})--(\ref{SRL}) one can easily see that
this procedure allows to extract $\sigma_{\rm LL}$, $\sigma_{\rm RR}$,
$\sigma_{\rm LR}$ and $\sigma_{\rm RL}$ by the four independent
measurements of
$\sigma_1(z^*_\pm)$ and $\sigma_2(z^*_\pm)$ at $P_e=\pm P$.

The kind of analysis given above shows that the separation of the helicity
cross sections can be performed by means of two independent integrated
observables (i.e., cross sections) at two different
values of polarization.\footnote{An alternative possibility to disentangle
the helicity cross sections in the process (\ref{proc}) based on differential
distributions was studied in \cite{Frere}. However, with limited statistics,
the approach exploiting integrated observables has an advantage.} Clearly, 
the measurement of four independent observables is a minimum to perform such 
separations. 

In the sequel, we shall make a comparison of the sensitivity to
contact-interaction couplings of the helicity cross sections determined by 
the procedure outlined above, and the corresponding discovery limits on  
$\Lambda_{\alpha\beta}$, to the results of an analysis based on a $\chi^2$ fit 
to the set of `conventional' observables represented by $\sigma(P=0)$, the 
forward-backward asymmetry (also for $P=0$)
\begin{equation}
\label{afb}
A_{\rm FB}=\frac{\sigma_{\rm FB}}{\sigma},
\end{equation}
the left-right asymmetry
\begin{equation}
A_{\rm LR}=\frac{\sigma_{\rm L}-\sigma_{\rm R}}{\sigma_{\rm L}+\sigma_{\rm R}}
=\frac{\sigma_{\rm LL}-\sigma_{\rm RR}+\sigma_{\rm LR}-
\sigma_{\rm RL}}{\sigma_{\rm LL}+\sigma_{\rm RR}+\sigma_{\rm LR}
+\sigma_{\rm RL}},
\label{alr}
\end{equation}
and the combined left-right forward-backward asymmetry 
\begin{equation}
A_{\rm LR,FB}=\frac{(\sigma_{\rm L}^{\rm F}-\sigma_{\rm R}^{\rm F})-
(\sigma_{\rm L}^{\rm B}-\sigma_{\rm R}^{\rm B})}
{(\sigma_{\rm L}^{\rm F}+\sigma_{\rm R}^{\rm F})+
(\sigma_{\rm L}^{\rm B}+\sigma_{\rm R}^{\rm B})}=
\frac{3}{4}\frac{\sigma_{\rm LL}-\sigma_{\rm RR}+\sigma_{\rm RL}-
\sigma_{\rm LR}}{\sigma_{\rm LL}+\sigma_{\rm RR}+\sigma_{\rm RL}+
\sigma_{\rm LR}},
\label{afbpol}
\end{equation}
where $\sigma_{\rm L}$ and $\sigma_{\rm R}$ denote the cross sections with
left-handed and right-handed electrons and unpolarized positrons.
%%%%%%%%%%%%%%%%%%%%%%%%%%%%%%%%%%%%%%%%%%%%%%%%%%%%%%%%%%%%%%%%%%%%%%%%
%\section{Optimization and radiative corrections}
\section{Generalization and radiative corrections}
%%%%%%%%%%%%%%%%%%%%%%%%%%%%%%%%%%%%%%%%%%%%%%%%%%%%%%%%%%%%%%%%%%%%%%%%
This extraction of helicity cross sections can be obtained more generally.
Indeed, let us divide the full angular range,
$\vert\cos\theta\vert\le 1$ into two parts, ($-1,\ z^*$) and
($z^*,\ 1$), with arbitrary $z^*$, and define two integrated cross sections as
\begin{eqnarray}
\label{sigma1}
\sigma_1(z^*)
&\equiv&\int_{z^*}^1\frac{\dd\sigma}{\dd\cos\theta}\dd\cos\theta
=\frac{1}{8}\left\{\left[8-(1+z^*)^3\right]\sigma_++(1-z^*)^3
\sigma_-\right\}, \\
\label{sigma2}
\sigma_2(z^*)
&\equiv&\int^{z^*}_{-1}\frac{\dd\sigma}{\dd\cos\theta}\dd\cos\theta
=\frac{1}{8}\left\{(1+z^*)^3\sigma_++
\left[8-(1-z^*)^3\right]\sigma_-\right\}.
\end{eqnarray}
Solving these two equations, one finds the general relations
\begin{eqnarray}
\label{sigmap}
\sigma_+
&=&\frac{1}{6(1-{z^*}^2)}\left[\left(8-(1-z^*)^3\right)
\sigma_1(z^*)-(1-z^*)^3\sigma_2(z^*)\right], \\
\label{sigmam}
\sigma_-
&=&\frac{1}{6(1-{z^*}^2)}\left[-(1+z^*)^3\sigma_1(z^*)
+\left(8-(1+z^*)^3\right)\sigma_2(z^*)\right],
\end{eqnarray}
that allow to disentangle the
helicity cross sections, using (\ref{SLL})--(\ref{SRL}) and
the availability of polarized beams.

In order to extract the helicity cross sections, we thus make use
of {\it two} sets of integrated cross sections.
The basic ones are $\sigma_1(z^*,P)$ and $\sigma_2(z^*,P)$,
which depend both on the kinematical cut $z^*$ and the polarization.
From these, as a second step, we construct the cross sections
$\sigma_+(P)$ and $\sigma_-(P)$ of the previous section,
which finally yield the helicity cross sections $\sigma_{\alpha\beta}$.

It is instructive to consider some particular cases depending on
the choice of $z^*$:

\par\noindent({\it i}) If one chooses $z^*=0$, then 
$\sigma_{1,2}=\sigma_{\rm F,B}$ and, from (\ref{sigmap}) and 
(\ref{sigmam}), $(7\sigma_{\rm F,B}-\sigma_{\rm B,F})/6=\sigma_\pm$.
%\begin{equation}
%\label{relation}
%{\sigma}_{\pm}=
%\frac{7}{6}\sigma_{\rm F,B}-
%\frac{1}{6}\sigma_{\rm B,F}=
%\frac{1}{2}\,\sigma \left(1\pm\frac{4}{3}A_{\rm FB}\right),
%\end{equation}
%where the total cross section is $\sigma=\sigma_{\rm F}+\sigma_{\rm B}$.
%\\[0.3cm]
\par\noindent({\it ii}) Requiring $z^*=z^*_+=1-2^{2/3}=-0.587$, one re-obtains 
the first relation in Eq.~(\ref{tildesigma+-}) and  
$(1-\gamma^{-1})\sigma_1+(1+\gamma^{-1})\sigma_2=\sigma_-$.
%\begin{equation}
%\label{iim}
%\sigma_-=-(\gamma^{-1} -1)\sigma_1+(\gamma^{-1} +1)\sigma_2.
%\end{equation}
%\\[0.1cm]
\par\noindent ({\it iii}) Taking $z^*=z^*_-=-1+2^{2/3}=0.587$, one re-obtains 
the second relation in Eq.~(\ref{tildesigma+-}) and 
$(1+\gamma^{-1})\sigma_1+(1-\gamma^{-1})\sigma_2=\sigma_+$.
%\begin{equation}
%\sigma_+=(\gamma^{-1} +1)\sigma_1-(\gamma^{-1} -1)\sigma_2.
%\end{equation}
%\\[0.1cm]
\par\noindent({\it iv}) For unpolarized initial beams ($P=0$), which is the 
case at LEP2, only linear combinations of helicity cross sections can be 
separated:
\begin{equation}
\label{unpol}
\frac{1}{4}(\sigma_{\rm LL}+\sigma_{\rm RR})=\sigma_+,\qquad 
\frac{1}{4}(\sigma_{\rm LR}+\sigma_{\rm RL})=\sigma_-.
\end{equation}
%\begin{eqnarray}
%\label{unpp}
%\frac{1}{4}(\sigma_{\rm LL}+\sigma_{\rm RR})&=&\sigma_+, \\
%\label{unpm}
%\frac{1}{4}(\sigma_{\rm LR}+\sigma_{\rm RL})&=&\sigma_-.
%\end{eqnarray}

The previous formulae continue to hold,
with the inclusion of one-loop SM electroweak radiative corrections, in the
form of improved Born amplitudes. Basically, the parameterization that
uses the
best known SM parameters $G_{\rm F}$, $M_Z$ and $\alpha(M^2_Z)$ is obtained by
replacing $\alpha\Rightarrow\alpha(M_Z^2)$ and  
$\sin^2\theta_W\Rightarrow\sin^2\theta^{\rm eff}_W$ in the above
equations and \cite{Hollik, Altarelli2}:
\begin{equation}
\label{impborn}
g^f_{\rm L}\Rightarrow
\frac{2}{\sqrt{\kappa}}
\left(I^f_{3L}-Q_f\sin^2\theta^{\rm eff}_W\right),\ 
g^f_{\rm R}\Rightarrow
-\frac{2 Q_f}{\sqrt{\kappa}}\sin^2\theta^{\rm eff}_W,\ 
\sin^2(2\theta^{\rm eff}_W)
\equiv\kappa=\frac{4\pi\alpha(M_Z^2)}{\sqrt{2}G_{\rm F}\, M_Z^2\rho},
\end{equation}
%\rightline{impborn}
with $\rho\approx 1+3 G_{\rm F} m^2_{\rm top}/(8\pi^2\sqrt{2})$.
Moreover, for the $Z$-propagator:
$\chi_Z(s)\Rightarrow s/(s-M_Z^2+i(s/M_Z^2)M_Z\Gamma_Z)$.

We take radiative corrections into account by means of the program
ZFITTER \cite{zfitter}, which has to be used along with ZEFIT,
adapted to the present discussion.
We thus include initial- and final-state radiation, and
the interference between them.
Initial state radiation (ISR)
is of major importance for contact interaction searches.
The observed cross section is significantly distorted in shape
and magnitude by the emission of real photons by the incoming
electrons and positrons \cite{Djouadi}.
In the mentioned program,
the hard photon radiation is calculated up to order $\alpha^2$
and the leading soft and virtual
corrections are summed to all orders by the exponentiation technique.
The final expression for the differential cross section is
of the same form as Eq.~(\ref{cross}),
where the scattering angle refers to that between the
final-state fermion $f$ and the $e^-$ beam
direction in the $f \bar f$ centre-of-mass system \cite{Was}.
The symmetric and antisymmetric parts of the cross section,
$\sigma_{s,a}$, are given by convolutions with the `radiators':
\begin{equation}
\label{coeff}
\sigma_s=\int\limits_{0}^{\Delta}
\dd k\, R^e_{\rm T}(k)\, \sigma(s'), \qquad
\sigma_a=\frac{4}{3}\int\limits_{0}^{\Delta}
\dd k\, R^e_{\rm FB}(k)\, \sigma_{\rm FB}(s'),
\end{equation}
with $s'=s(1-k)$, and $k=E_\gamma/E_{\rm beam}$ the fraction of energy
lost by radiation,
and $R^e_{\rm T}(k)$ and $R^e_{\rm FB}(k)$ the radiator functions,
whose explicit expression can be found in ref.~\cite{zfitter}.

Due to the radiative return to the $Z$ resonance at $\sqrt{s}>M_Z$, the
energy spectrum of the radiated photons is peaked around
$k_{\rm peak}\approx 1-M^2_Z/s$ \cite{Djouadi}.
In order to increase the signal originating from contact interactions,
events with hard photons should be eliminated
by an appropriate cut $\Delta<k_{\rm peak}$ on the photon energy.
Since the form of the corrected cross section
is the same as that of Eq.~(\ref{cross}), it follows that
the radiatively-corrected $\sigma_{1,2}$ can also be defined
by Eqs.~(\ref{sigma1}) and (\ref{sigma2}),
with the same value for $z^*$.
For our numerical analysis, we use $m_{\rm top}=175$~GeV, $m_H=100$~GeV and 
a cut $\sqrt{s^\prime}\ge 0.9\sqrt{s}$ to avoid the radiative return to 
the $Z$ peak for $\sqrt{s}=0.5$ TeV.

The convenience of $\sigma_{1,2}$ of Eqs.~(\ref{sigma1}) and (\ref{sigma2})
as experimental integrated observables will be appreciated in the next
Section, where the numerical analysis and the corresponding bounds
on $\Lambda$'s will be `optimized' by suitable choices of the $z^*$
values that maximize the sensitivity to such non-standard interactions.
%%%%%%%%%%%%%%%%%%%%%%%%%%%%%%%%%%%%%%%%%%%%%%%%%%%%%%%%%%%%%%%%%%%%%%%%
\section{Sensitivity and optimization}
%%%%%%%%%%%%%%%%%%%%%%%%%%%%%%%%%%%%%%%%%%%%%%%%%%%%%%%%%%%%%%%%%%%%%%%%
In the case where no deviation from the SM is observed, one can make an
assessment of the sensitivity of the process (\ref{proc}) to the contact 
interaction parameters, based on the expected experimental accuracy on the 
observables $\sigma_{\alpha\beta}$. Such sensitivity numerically determines 
the bounds on the contact-interaction scales $\Lambda_{\alpha\beta}$ 
that can be derived from the experimental data and, basically, 
is determined by the comparison
of deviations from the SM predictions due to the contact-interaction terms 
with the attainable experimental uncertainty. Accordingly, we define the 
`significance' of each helicity cross section by the ratio:
\begin{equation}
\label{signif}
{\cal S}
=\frac{|\Delta\sigma_{\alpha\beta}|}{\delta\sigma_{\alpha\beta}}, 
\end{equation}
where $\Delta\sigma_{\alpha\beta}$ is the deviation from the SM prediction, 
dominated for $\sqrt s\ll \Lambda_{\alpha\beta}$ by the interference term: 
\begin{equation}
\Delta\sigma_{\alpha\beta}\equiv
\sigma_{\alpha\beta}-\sigma_{\alpha\beta}^{\rm SM}\simeq
2 N_C\, \sigma_{\rm pt}
\left(Q_e\, Q_f+g_{\alpha}^e\, g_{\beta}^f\,\chi_Z\right)
\frac{s\eta_{\alpha\beta}}{\alpha\Lambda_{\alpha\beta}^2},
\label{deltasig}
\end{equation}
and $\delta\sigma_{\alpha\beta}$ is the expected experimental uncertainty on 
$\sigma_{\alpha\beta}$, combining statistical and systematic uncertainties.

For example, adding uncertainties in quadrature, the uncertainty on 
$\sigma_{\rm LL}$, indirectly measured {\it via} $\sigma_1$
and $\sigma_2$ (see Eqs.~(\ref{SLL}) and (\ref{sigmap})), 
is given by
\begin{eqnarray}
\label{uncet}
\left(\delta \sigma_{LL}\right)^2
=a^2(z^*)\left ( \frac{1+P}{P} \right )^2 (\delta\sigma_1 (z^*,-P))^2
+a^2(z^*) \left ( \frac{1-P}{P} \right )^2 (\delta\sigma_1 (z^*,P))^2
\nonumber \\
+b^2(z^*) \left ( \frac{1+P}{P} \right )^2 (\delta\sigma_2 (z^*,-P))^2
+b^2(z^*) \left ( \frac{1-P}{P} \right )^2 (\delta\sigma_2 (z^*,P))^2, 
\end{eqnarray}
where
\begin{equation}
a(z^*)=\frac{8-(1-z^*)^3}{6(1-{z^*}^2)}, \qquad
b(z^*)=-\frac{(1-z^*)^3}{6(1-{z^*}^2)} .
\end{equation}
Analogous expressions hold for the combinations related to the uncertainties
$\delta\sigma_{\rm RR}$, $\delta\sigma_{\rm LR}$ and $\delta\sigma_{\rm RL}$.
Numerically, in the situation of small deviations from the SM we are 
considering, we can use to a very good approximation the SM predictions for 
the cross sections $\sigma_{1,2}$ to assess the expected $\delta\sigma_{1,2}$ 
and therefore of the uncertainties $\delta\sigma_{\alpha\beta}$ in the 
denominator of (\ref{signif}). Basically, the directly measured integrated 
cross sections $\sigma_{1,2}$ of Eqs.~(\ref{sigma1}) and (\ref{sigma2}) and, 
correspondingly, the uncertainties $\delta\sigma_{\alpha\beta}$, are dependent 
on the value of $z^*$, which can be considered in general as an input 
parameter related to given experimental conditions (see, {\it e.g.}, 
Eq.~(\ref{uncet})). Since the deviation 
$\Delta\sigma_{\alpha\beta}$ of Eq.~(\ref{deltasig}) is independent of $z^*$, 
the full sensitivity of a given helicity cross section to the relevant 
contact-interaction parameter is determined by the corresponding size and 
$z^*$ behavior of the uncertainty $\delta\sigma_{\alpha\beta}$. Then, the 
optimization would be obtained by choosing for $z^*$ the value $z^*_{\rm opt}$ 
where the uncertainty $\delta\sigma_{\alpha\beta}$ becomes minimum, 
{\it i.e.}, where the corresponding sensitivity Eq.~(\ref{signif}) has a 
maximum. As anticipated, we estimate the required $z^*$ behavior from the 
known SM cross sections. 

Accordingly, statistical uncertainties will be given by 
\begin{equation}
\label{delsi}
(\delta\sigma_i)^2_{\rm stat}\simeq(\delta\sigma_i^{\rm SM})^2_{\rm stat}
=\frac{\sigma_i^{\rm SM}}{\epsilon\, \Lumint}, \qquad i=1,2, 
\end{equation}
where $\Lumint$ is the integrated luminosity,
and $\epsilon$ is the efficiency for detecting the final state under
consideration.
The equation that determines the values of $z^*$ that minimize the
statistical uncertainties on $\sigma_{\alpha\beta}$ is:
\begin{equation}
z^*=-3 \frac{1-r_{\alpha\beta}}{1+r_{\alpha\beta}}
\frac{{z^*}^4 -6{z^*}^2 -3}{{z^*}^4 -2{z^*}^2 -23},
\label{zopt}
\end{equation}
where
\begin{equation}
\label{rL}
r_{\rm LL}=r_{\rm LR}=\frac{(1+3P^2)\sigma_{\rm LR}^{SM}
+(1-P^2)\sigma_{\rm RL}^{SM}}
{(1+3P^2)\sigma_{\rm LL}^{SM}+(1-P^2)\sigma_{\rm RR}^{SM}},
\end{equation}
%\begin{equation}
%\label{rR}
%r_{\rm RR}=r_{\rm RL}=\frac{(1+3P^2)\sigma_{\rm RL}^{SM}
%+(1-P^2)\sigma_{\rm LR}^{SM}}
%{(1+3P^2)\sigma_{\rm RR}^{SM}+(1-P^2)\sigma_{\rm LL}^{SM}}.
%\end{equation}
and $r_{\rm RR}=r_{\rm RL}$ is obtained by replacing 
${\rm L}\leftrightarrow{\rm R}$ in (\ref{rL}). 
As one can see, the location of $z^*$ that
minimizes the statistical uncertainty, only depends on the SM parameters
and $P$, and is independent from the luminosity and efficiency of 
reconstruction $\epsilon$ of a final-state fermion. 
In a left-right symmetric theory, the above ratios $r_{\alpha\beta}$
would all be 1, and in this case $z^*=0$.
However, in the SM, depending on flavour and energy, $r_{\alpha\beta}$
may be less than, or larger than unity.
Since the $z^*$-dependent fraction in (\ref{zopt}) is positive
for $z^{*2}\le1$, it follows that the solutions satisfy $z^*<0$
if $r_{\alpha\beta}<1$ and {\it vice versa}.
We also note that the location is the same for the $\rm LL$ and $\rm LR$ 
configurations, and likewise for $\rm RR$ and $\rm RL$, while numerically 
the sensitivities are different. 
The numerical solutions of Eq.~(\ref{zopt}) are reported in Table 1 
for different values of longitudinal polarization of electrons $P$.

\begin{table}[t]
\centering
\begin{tabular}{|c|c|c|c||c|c|}
\hline
process
& $P$ & $r_{LL}$; $r_{LR}$ & $z^*_{\rm opt}$ & $r_{RR}$; $r_{RL}$ &
                                  $z^*_{\rm opt}$ \\ \hline & 1.0  & 0.19
                                  & -0.32              & 0.24   &   -0.28
\\ \cline{2-6}
$e^+e^- \to l^+l^-$  &  0.9  &  0.19 & -0.32              &
0.23   &   -0.28             \\ \cline{2-6}
&  0.5  & 0.20  &  -0.30
                                   &  0.22  &   -0.29            \\
                                   \hline \hline
& 1.0  & 0.06  & -0.51
                                   & 0.30   &   -0.23              \\
\cline{2-6}
$e^+e^- \to b \bar b$  &  0.9  &  0.07 & -0.50              &
0.24   &   -0.27             \\ \cline{2-6}
&  0.5  & 0.08  &  -0.47
                                   &  0.13  &   -0.32            \\
                                  \hline \hline
& 1.0  & 0.14  & -0.38
                                  & 0.07   &   -0.48              \\
\cline{2-6}
$e^+e^- \to c \bar c$  &  0.9  &  0.13 & -0.38              &
0.08   &   -0.47             \\ \cline{2-6}
&  0.5  & 0.12  &  -0.39
                                   &  0.10  &   -0.46            \\
\hline
\end{tabular}
\caption
{Optimal values of  $z^{*}_{\rm opt}$ obtained from
Eq.~(\ref{zopt}) at $E_{\rm c.m.}=0.5$~TeV.}
\label{tab:tab1}
\end{table}

Clearly, the optimal values of $z^*$ reported in Table~1 can be applied in
practice in the case of low statistics, where the statistical uncertainty
dominates over the systematic one. In the general case, the latter
can affect the determination of the value of $z^*_{\rm opt}$, especially
in the case of higher luminosity, where it may dominate over the
statistical uncertainties. Combining, again in quadrature, statistical and
systematic uncertainties on $\sigma_{1,2}$, we have:
\begin{equation}
\label{delsi1}
(\delta\sigma_i)^2\simeq(\delta\sigma_i^{\rm SM})^2=
\frac{\sigma_i^{\rm SM}}{\epsilon\, \Lumint}
+\left(\delta^{\rm sys}\sigma_i^{\rm SM}\right)^2.
\end{equation}
%\par 

Numerically, for $\sigma_{1,2}$ we take into account the expected 
identification efficiencies, $\epsilon$ \cite{Damerell} and the systematic 
uncertainties, $\delta^{\rm sys}$, on the various fermionic final states, for 
which we assume: for leptons: $\epsilon=95\%$ and $\delta^{\rm sys}=0.5\%$;
for $b$ quarks: $\epsilon=60\%$ and $\delta^{\rm sys}=1\%$;
for $c$ quarks: $\epsilon=35\%$ and $\delta^{\rm sys}=1.5\%$. As
concerns the systematic uncertainty, we assume the same $\delta^{\rm sys}$ 
for $i=1,2$, and independent of $z^*$ in the relevant angular range.   

We consider the LC with the following options: $\sqrt{s}=0.5$~TeV,
$\Lumint=50\ \mbox{fb}^{-1}$ up to $\Lumint=500\ \mbox{fb}^{-1}$ to assess
the 
role of statistics, $P=0.9$ and $|\cos\theta|\le 0.99$.  We assume half the 
total integrated luminosity quoted above for both values of the electron 
polarization, $P_e=\pm P$.

%%%%%%%%%%%%%%%%%%%%%%%%%%%%%%%%%%%%%%%%%%%%%%%%%%%%%%%%%%%%%%%%%%%%%%%%
\begin{figure}
\begin{center}
\setlength{\unitlength}{1cm}
\begin{picture}(7.0,6.5)
\put(-1.5,-0.5){
\mbox{\epsfysize=9cm\epsffile{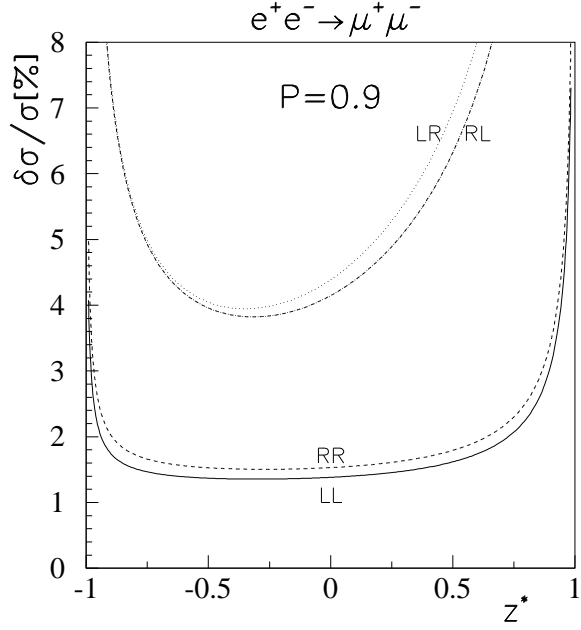}}}
\end{picture}
%\vspace*{-4mm}
\caption{
The uncertainty on the helicity cross sections $\sigma_{\alpha\beta}$ 
in the SM
as a function of $z^*$ for the process $e^+e^-\to\mu^+\mu^-$ at 
$\sqrt{s}=0.5$~TeV, $\Lumint=50\ \mbox{fb}^{-1}$, $P=0.9$, 
$\epsilon=95\%$ and
$\delta^{\rm sys}=0.5\%$. Radiative corrections are included.}
\end{center}
\end{figure}
%%%%%%%%%%%%%%%%%%%%%%%%%%%%%%%%%%%%%%%%%%%%%%%%%%%%%%%%%%%%%%%%%%%%%%%%

%%%%%%%%%%%%%%%%%%%%%%%%%%%%%%%%%%%%%%%%%%%%%%%%%%%%%%%%%%%%%%%%%%%%%%%%
\begin{figure}
\begin{center}
\setlength{\unitlength}{1cm}
\begin{picture}(7.0,6.5)
\put(-5.5,-0.5){
\mbox{\epsfysize=9cm\epsffile{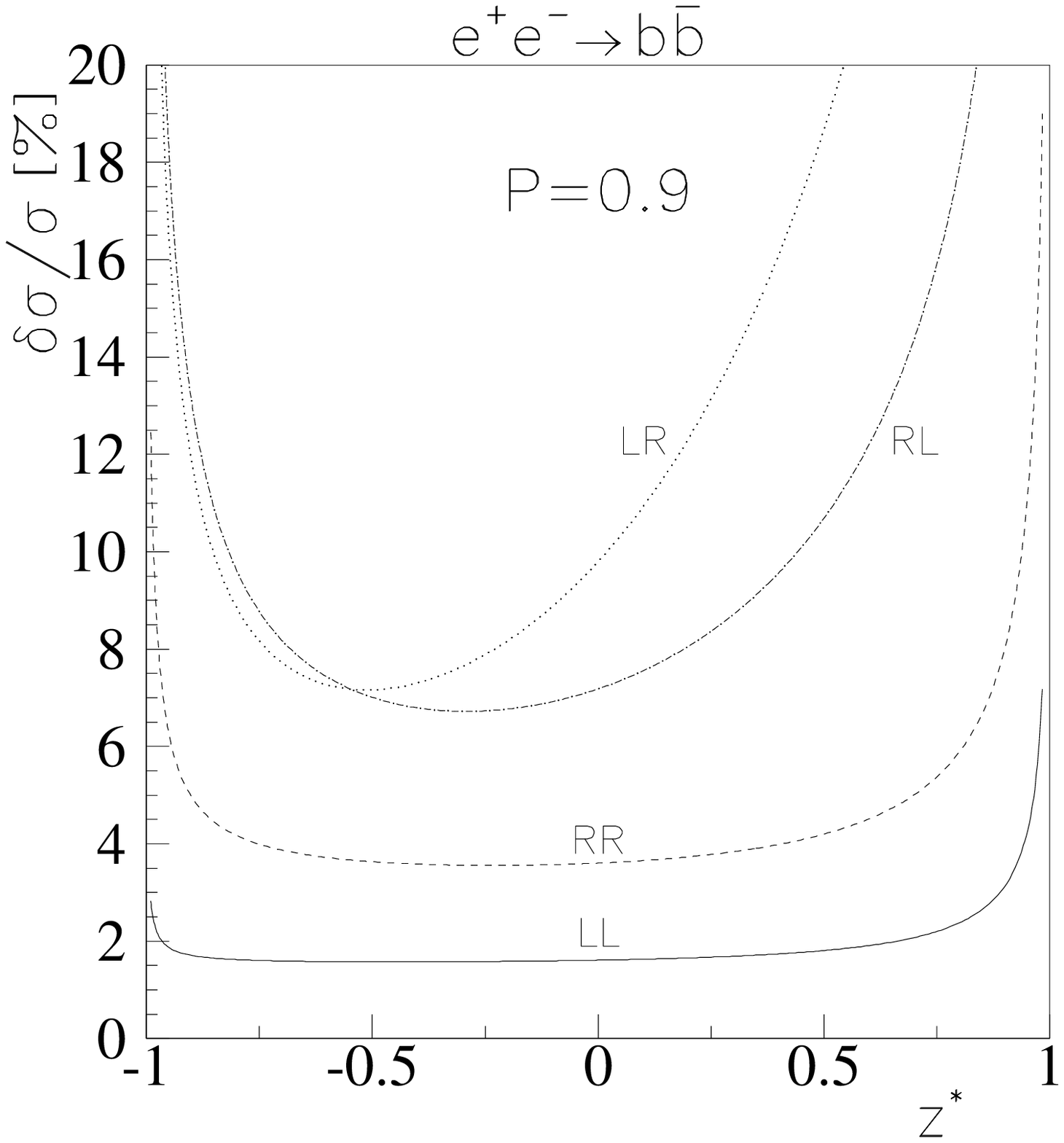}}
\mbox{\epsfysize=9cm\epsffile{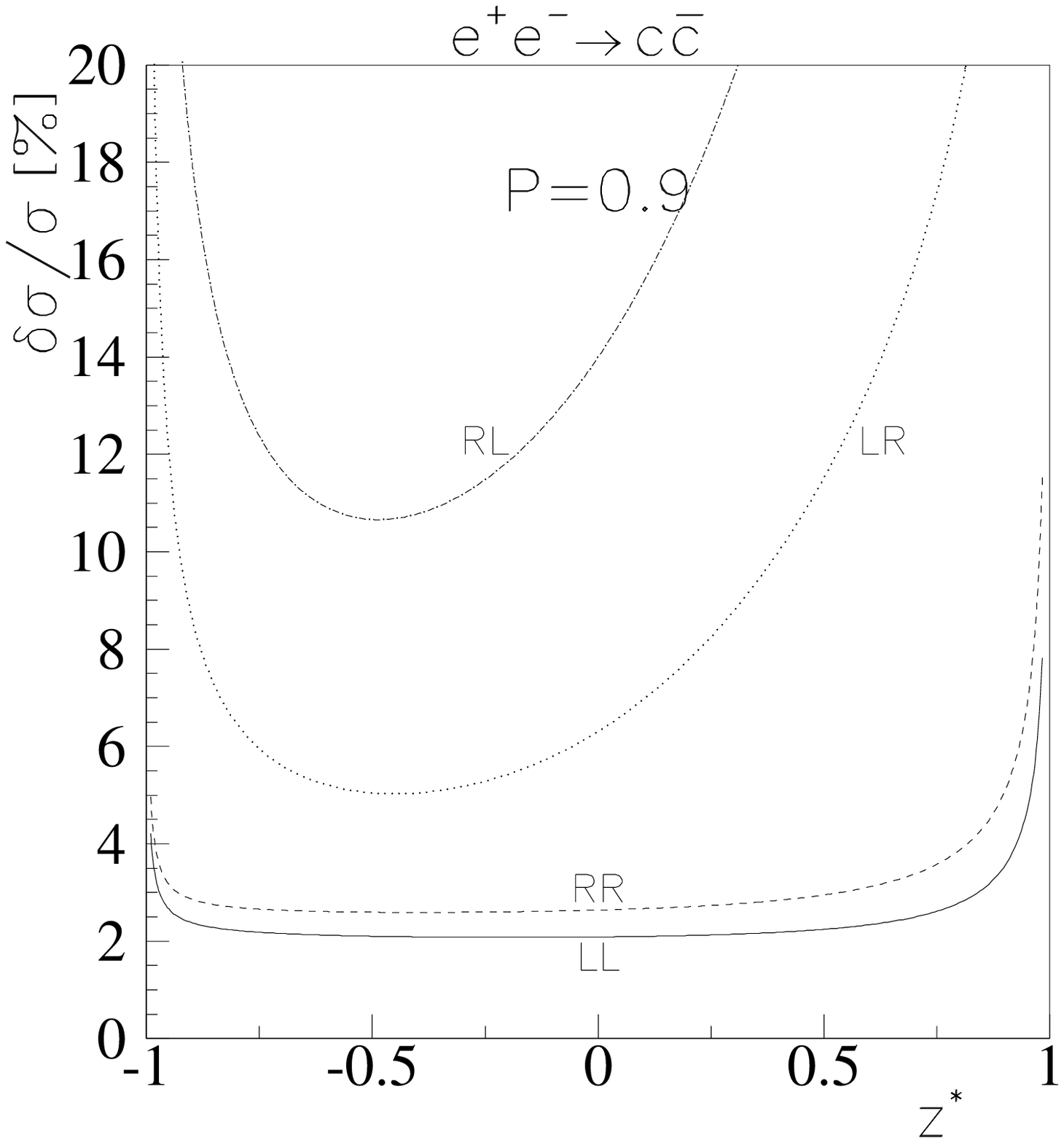}}}
\end{picture}
\caption{
Same as in Fig.~1, but for the processes $e^+e^-\to b\bar{b}$
at $\epsilon=60\%$ and $\delta^{\rm sys}=1.0\%$, and
$e^+e^-\to c\bar{c}$
at $\epsilon=35\%$ and $\delta^{\rm sys}=1.5\%$.}
\end{center}
\end{figure}
%%%%%%%%%%%%%%%%%%%%%%%%%%%%%%%%%%%%%%%%%%%%%%%%%%%%%%%%%%%%%%%%%%%%%%%%

The relative uncertainties, $\delta\sigma_{\alpha\beta}/\sigma_{\alpha\beta}$,
on the helicity cross sections, are shown as functions of $z^*$
in Figs.~1,2 for different final states.
The optimal kinematical parameters  $z^*_{\rm opt}$ where the sensitivity
is a maximum, can easily be obtained from these figures.
Inclusion of systematic errors results in changing the location
of $z^*_{\rm opt}$ such that it increases for the cases of LL and RR 
helicity configurations and decreases for LR and RL.
When the two intervals are very different, $z^*\to\pm 1$, the uncertainty
blows up like $1/\sqrt{1\mp z^{*}}$, because the corresponding
cross section $\sigma_1$ or $\sigma_2$ vanishes with $1\mp z^*$, respectively.
Since, in these figures, the lowest value for $\Lumint$ is considered, the 
optimal values of $z^*$ are close to those in Table 1, because the
statistical uncertainty dominates in the considered situation. In Table~2,
we report the behaviour of $z^*_{\rm opt}$ with luminosity. 
%%%%%%%%%%%%%%%%%%%%%%%%%%%%%%%%%%%%%%%%%%%%%%%%%%%%%%%%%%%%%%%%%%%%%%%%
\begin{table}
{\small
\begin{center}
\begin{tabular}{|c||c|c|c||c|c|c||c|c|c||c|c|c|}
\hline
\multicolumn{1}{|c||}{$\Lumint$} &
\multicolumn{3}{|c||}{$z^{\ast}_{LL}$} &
\multicolumn{3}{|c||}{$z^{\ast}_{RR}$}& 
\multicolumn{3}{|c||}{$z^{\ast}_{LR}$}&
\multicolumn{3}{|c|}{$z^{\ast}_{RL}$} \\ 
\multicolumn{1}{|c||}{${\rm fb}^{-1}$} &
\multicolumn{1}{ c }{$l^+l^-$} &
\multicolumn{1}{ c }{$b\bar b$} &
\multicolumn{1}{ c||}{$c\bar c$} &
\multicolumn{1}{ c }{$l^+l^-$} &
\multicolumn{1}{ c }{$b\bar b$} &
\multicolumn{1}{ c||}{$c\bar c$} &
\multicolumn{1}{ c }{$l^+l^-$} &
\multicolumn{1}{ c }{$b\bar b$} &
\multicolumn{1}{ c||}{$c\bar c$} &
\multicolumn{1}{ c }{$l^+l^-$} &
\multicolumn{1}{ c }{$b\bar b$} &
\multicolumn{1}{ c |}{$c\bar c$}  \\ \hline
 50&-0.30&-0.45&-0.19 &-0.27&-0.23&-0.39 &-0.35&-0.51&-0.45& 
-0.32&-0.29&-0.49 \\ \hline
100&-0.28&-0.37&-0.09 &-0.24&-0.20&-0.34 &-0.36&-0.53&-0.50& 
-0.33&-0.31&-0.51 \\ \hline
200&-0.22&-0.33&0.13  &-0.19&-0.14&-0.23 &-0.39&-0.56&-0.55&
-0.35&-0.34&-0.54 \\ \hline
300&-0.17&-0.25&0.28  &-0.13&-0.08&-0.11 &-0.41&-0.58&-0.58&
-0.37&-0.37&-0.56 \\ \hline
400&-0.11&-0.17&0.35  &-0.09&-0.03&-0.00 &-0.43&-0.59&-0.60&
-0.40&-0.39&-0.58 \\ \hline
500&-0.07&-0.03&0.41  &-0.05&-0.01&0.11  &-0.44&-0.61&-0.62&
-0.41&-0.40&-0.59 \\ \hline
\end{tabular}
\end{center}  
\caption{
$z^*_{\rm opt}$ vs. ${\cal L}_{int}$ for the processes
$e^+e^- \to \mu^+\mu^-;\ b \bar b;\ c \bar c$ at
$E_{C.M.}=500$ GeV and $P=0.9$. Radiative corrections are included.
} 
} % end \small
\end{table}
%%%%%%%%%%%%%%%%%%%%%%%%%%%%%%%%%%%%%%%%%%%%%%%%%%%%%%%%%%%%%%%%%%%%%%%%

In order to assess the increase of sensitivity obtained by optimization,
one should compare the corresponding uncertainties at $z^*_{\rm opt}$ with 
those obtained without optimization, at $z^*_\pm$ of Eq.~(\ref{zpm}). 
Figs.~1,2 show that, in the LL and RR cases, optimization results in a
rather modest increase of sensitivity and of the corresponding discovery 
limits on $\Lambda_{\rm RR}$ and $\Lambda_{\rm LL}$ (by a few percent), 
since the $z^*$ behavior of the uncertainty is rather flat. 
Conversely, in the LR and RL cases optimization can substantially increase 
the sensitivity and the corresponding reachable lower bounds on 
$\Lambda_{\rm LR}$ and $\Lambda_{\rm RL}$ (up to a factor of about 2 
for the $c\bar c$ case).
%%%%%%%%%%%%%%%%%%%%%%%%%%%%%%%%%%%%%%%%%%%%%%%%%%%%%%%%%%%%%%%%%%%%%%%%
\section{Bounds on $\Lambda_{\alpha\beta}$ and conclusions}
%%%%%%%%%%%%%%%%%%%%%%%%%%%%%%%%%%%%%%%%%%%%%%%%%%%%%%%%%%%%%%%%%%%%%%%%
To obtain discovery limits on the four-fermion contact interaction, for
each helicity cross section we define a $\chi^2$ (see Eq.~(\ref{signif})):
\begin{equation}
\label{chisq}
\chi^2
=\left(\frac{\Delta\sigma_{\alpha\beta}}{\delta\sigma_{\alpha\beta}}\right)^2.
\end{equation}
As a criterion to constrain the allowed values of the contact interaction
parameters by the non-observation of the corresponding deviations within 
the expected uncertainty $\delta\sigma_{\alpha\beta}$, we impose 
$\chi^2<\chi^2_{\rm CL}$, where the actual value of $\chi^2_{\rm CL}$ 
specifies the desired `confidence' level.  
The deviations from the SM predictions $\Delta\sigma_{\alpha\beta}$
depend on a single `effective' non-standard parameter, namely, 
the product of the known relevant SM coupling and contact-interaction 
coupling in (\ref{deltasig}). 
Accordingly, in a $\chi^2$ analysis of data on $\sigma_{\alpha\beta}$, 
a fit in one effective parameter is involved.
Therefore, we take $\chi^2_{\rm CL}=3.84$ for 95\% C.L. with a 
one-parameter fit.

%%%%%%%%%%%%%%%%%%%%%%%%%%%%%%%%%%%%%%%%%%%%%%%%%%%%%%%%%%%%%%%%%%%%%%%%
\begin{figure}
\begin{center}
\setlength{\unitlength}{1cm}
\begin{picture}(8.0,8.0)
\put(-1.,-0.5){
\mbox{\epsfysize=10cm\epsffile{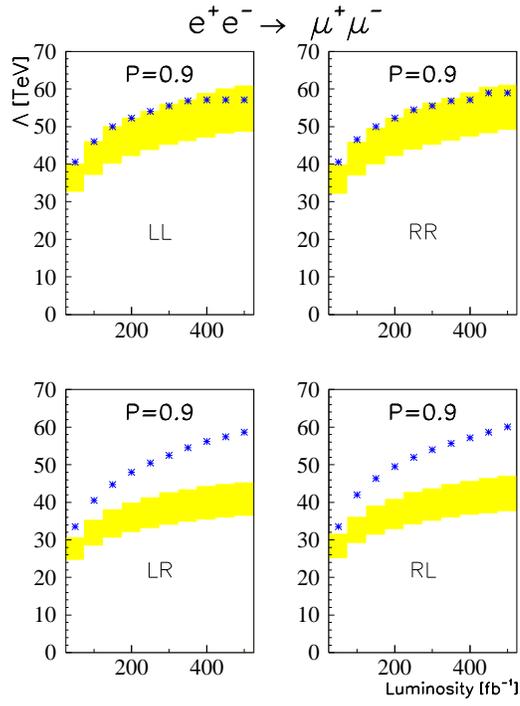}}}
\end{picture}
\caption{
Reach in $\Lambda$ at 95\% C.L., for the process $e^+e^-\to\mu^+\mu^-$
{\it vs.} integrated luminosity. Stars and shaded area indicate, respectively, 
constraints from helicity cross sections and from the set of observables 
$\sigma$, $A_{\rm FB}$, $A_{\rm LR}$ and $A_{\rm LR,FB}$. }
\end{center}
\end{figure}
%%%%%%%%%%%%%%%%%%%%%%%%%%%%%%%%%%%%%%%%%%%%%%%%%%%%%%%%%%%%%%%%%%%%%%%%

%%%%%%%%%%%%%%%%%%%%%%%%%%%%%%%%%%%%%%%%%%%%%%%%%%%%%%%%%%%%%%%%%%%%%%%%
\begin{figure}
\begin{center}
\setlength{\unitlength}{1cm}
\begin{picture}(8.0,8.0)
\put(-5.2,-0.5){
\mbox{\epsfysize=10cm\epsffile{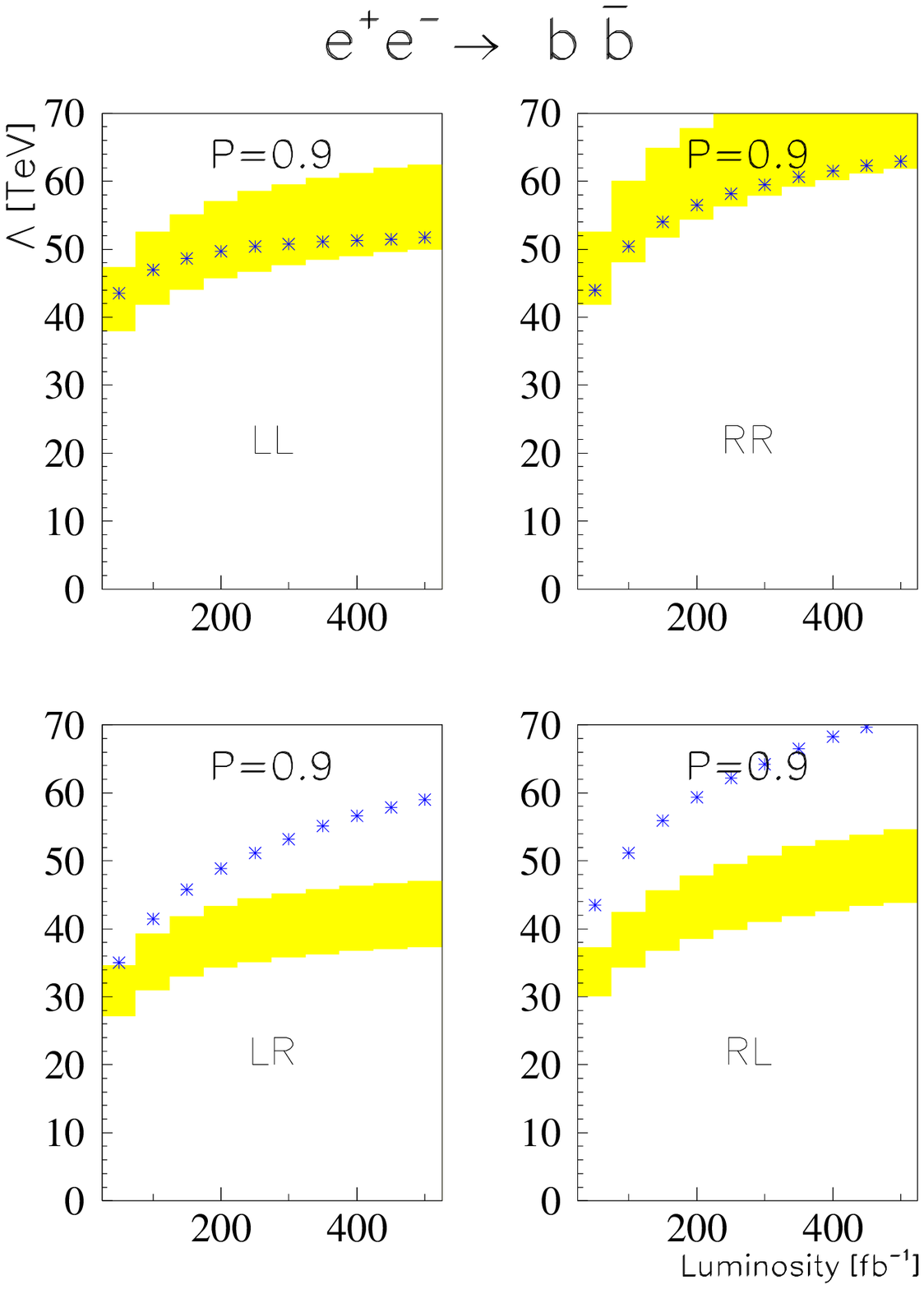}}
\mbox{\epsfysize=10cm\epsffile{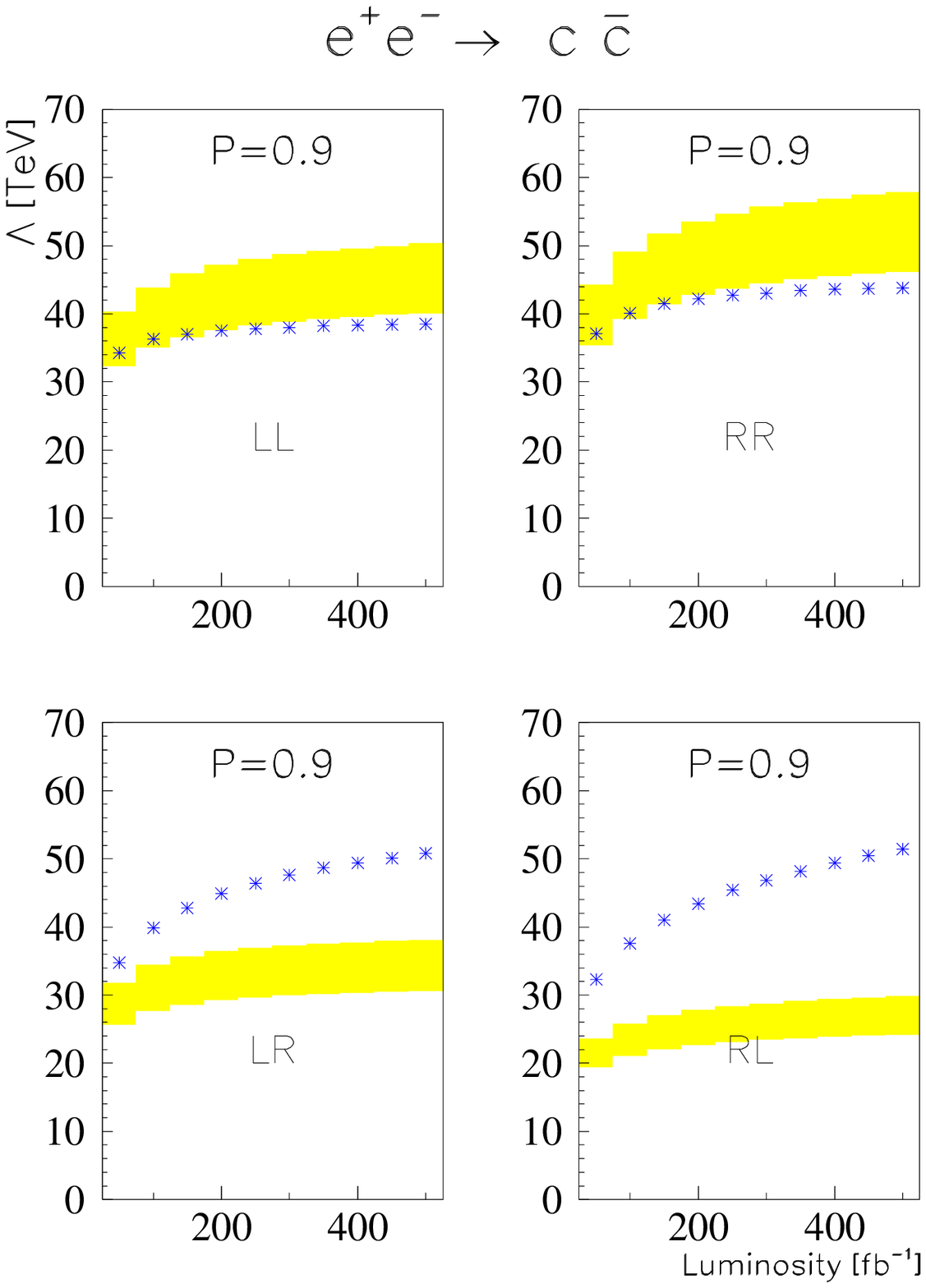}}}
\end{picture}
\caption{
Same as in Fig.~3, but for the processes $e^+e^-\to b \bar b$ and
$e^+e^-\to c \bar c$.}
\end{center}
\end{figure}
%%%%%%%%%%%%%%%%%%%%%%%%%%%%%%%%%%%%%%%%%%%%%%%%%%%%%%%%%%%%%%%%%%%%%%%%

The corresponding discovery reach for the mass scale parameters
$\Lambda_{\alpha\beta}$ as a function of luminosity $\Lumint$, obtained
from the determination of helicity cross sections {\it via} the measurement 
of $\sigma_1$ and $\sigma_2$ at $P=0.9$ and the optimal $z^*$ with the 
assumed values of the efficiencies $\epsilon$ and of the systematic 
uncertainties $\delta^{\rm sys}$, is represented by the stars
in Figs.~3,4. These figures show that the helicity cross
sections $\sigma_{\alpha\beta}$ are quite sensitive to contact
interactions, with discovery limits that, at the highest considered 
luminosity 500~$\mbox{fb}^{-1}$ (and $P=0.9$), can range from 80 up to 
140 times the c.m.\ energy, depending on the considered final fermion state. 
Indeed, the best sensitivity is achieved for the $b\bar{b}$ final state, 
while the worst one corresponds to the $c\bar{c}$ channel.
Moreover, we find that decreasing the electron polarization from
$P=1$ to $P=0.5$ results in a worsening of the sensitivity by $20-40\%$, 
depending on the final state. 

Regarding the role of the assumed uncertainties on the observables under
consideration, in the cases of $\Lambda_{\rm LR}$ and $\Lambda_{\rm RL}$
the uncertainty turns out to be numerically dominated by the statistical one,
especially for the smaller values of the luminosity, and the bounds have
less sensitivity to the value of the systematic uncertainty. Conversely, 
in the cases of $\Lambda_{\rm LL}$ and $\Lambda_{\rm RR}$ the results are 
more dependent on the assumed value of the systematic uncertainty. 
Asymptotically, for no systematic uncertainty, the bounds on 
$\Lambda_{\alpha\beta}$ scale like $\Lumint^{1/4}$ \cite{Zeppenfeld2}, 
which would give a factor 1.8 in improvement from 50 to 500 $\mbox{fb}^{-1}$.
Moreover, from Eqs.~(\ref{s+}) and (\ref{s-}), a further
improvement in the sensitivity to the various $\Lambda$-scales in  
Figs.~3,4 could be obtained if both $e^-$ and $e^+$ longitudinal
polarizations were available \cite{Accomando}.\footnote{It seems that
a significant positron polarization, of the order of 0.6, might be
achievable.  
We hope to return to a quantitative study, taking into account also
$e^+$ polarization, elsewhere.} Finally, regarding the role of radiative
corrections, one finds that the initial state radiation can lower 
the search reach by 15--20\%.

It is instructive to compare these bounds with those obtained from a
$\chi^2$ fit to the set of `conventional' observables ${\cal O}_i$=$\sigma$, 
$A_{\rm FB}$, $A_{\rm LR}$ and $A_{\rm LR,FB}$ quoted in Sect.~2. 
In this case, the relevant $\chi^2$ can be constructed as follows:
\begin{equation}
\label{chi2} 
\chi^2
=\sum_i
\left(\frac{{\cal O}_i-{\cal O}_i^{\rm SM}}{\delta{\cal O}_i^{\rm SM}}
\right)^2,
\end{equation}
where the sum is over observables included in the $\chi^2$, and
$\delta{\cal O}_i^{SM}$ is the expected experimental uncertainty on 
the observable ${\cal O}_i$ evaluated, as before, by using the known SM 
cross sections.
Clearly, in contrast to the case of helicity cross sections, the $\chi^2$ of
Eq.~(\ref{chi2}) simultaneously depends on all four contact interaction 
couplings. 
Thus, potential cancellations among those, 
{\it a priori} free, parameters can occur and lead to either looser, 
or correlated, bounds on $\Lambda_{\alpha\beta}$. To be quantitative, we may
consider the following two representative cases.

The first case represents the simplest situation where all
observables ${\cal O}_i$ depend on only one contact interaction parameter. 
This can be realized, {\it e.g.}, within some specific model. 
In this case, one can reduce it to a one-parameter fit, but the corresponding 
bounds on the individual mass scale parameters $\Lambda_{\alpha\beta}$ 
refer to the specific model. These bounds can be obtained from 
the $\chi^2$ of Eq.~(\ref{chi2}) by a procedure analogous to that 
described above for the helicity cross sections, with $\chi^2_{\rm CL}=3.84$. 
For that one-parameter case the 95\% C.L. bounds on $\Lambda_{\alpha\beta}$
are represented by the upper contours of the shaded regions in Figs.~3,4.

The second case assumes a full four-parameter dependence of the
observables. The model-independent analysis including the complete set of 
four contact interaction couplings as free parameters would identify 
the observability domain as the region in four-dimensional space external 
to the surface determined by the equation
$\chi^2= \chi^2_{\rm CL}$ (where, now, $\chi^2_{\rm CL}=9.49$ corresponds
to 95\% C.L.). In order to get the bounds on an individual parameter, 
one should project this surface onto the axis related to that parameter, 
an analysis that is not simple and, actually, is outside the scope of 
the present paper. 
We consider here a simplified, but somewhat extreme, case which by 
construction excludes any accidental cancellation among contributions 
induced by different parameters.
It corresponds to the projection of the above-mentioned surface onto the
axis related to one contact parameter, 
taking all the others at zero value. 
The corresponding bounds on the individual $\Lambda_{\alpha\beta}$ are
represented by the lower contours of the shaded areas in Figs.~3,4. 
However, in general such contours may be substantially lowered by 
the potential cancellations, whose effects can spoil the sensitivity 
to the point that, in some extreme case, a parameter might even remain
unconstrained. Therefore, the lower contours in Figs.~3,4 just indicate the 
maximal potential sensitivity to contact-interaction couplings that can be 
expected from the model-independent $\chi^2$ analysis of the `conventional' 
observables in the ideal situation of no cancellation.

The comparison of bounds displayed in Figs.~3,4 from `conventional'
observables with those obtained from the analysis using helicity 
cross sections as basic observables shows that for the LL and RR cases 
the two kinds of analysis are numerically comparable.  
Instead, for the LR and RL combinations the sensitivity of helicity 
cross sections to the relevant contact interaction parameters is
appreciably higher. This is mostly due to the optimization through the
dependence on the kinematical parameter $z^*$ described in Sect.~3. 
One should also emphasize that, besides the high sensitivity, helicity 
cross sections have the qualitative advantage of providing, by definition,
unambiguous and fully model-independent information on the new physics
parameters of Eq.~(\ref{lagra}).

%%%%%%%%%%%%%%%%%%%%%%%%%%%%%%%%%%%%%%%%%%%%%%%%%%%%%%%%%%%%%%%%%%%%%%%%

%\newpage
%%%%%%%%%%%%%%%%%%%%%%%%%%%%%%%%%%%%%%%%%%%%%%%%%%%%%%%%%%%%%%%%%%%%%%%%
%\section*{Acknowledgements}
\medskip
\leftline{\bf Acknowledgements}
\par\noindent
It is a pleasure to thank Arnd Leike, Sabine Riemann and Luca Trentadue 
for helpful discussions.
This research has been supported by the Research Council of Norway,
and by MURST (Italian Ministry of University, Scientific Research
and Technology).
%%%%%%%%%%%%%%%%%%%%%%%%%%%%%%%%%%%%%%%%%%%%%%%%%%%%%%%%%%%%%%%%%%%%%%%%

%\newpage

%%%%%%%%%%%%%%%%%%%%%%%%%%%%%%%%%%%%%%%%%%%%%%%%%%%%%%%%%%%%%%%%%%%%%%%%

\end{document}